\documentstyle[aps,prb,twocolumn,floats,epsf]{revtex}

\begin{document}
\draft

\twocolumn[\hsize\textwidth\columnwidth\hsize\csname @twocolumnfalse\endcsname

\title{\bf Local-moment formation in gapless Fermi systems}
\bigskip
\author{Carlos Gonzalez-Buxton
and Kevin Ingersent\cite{byline}}
\address{Department of Physics, University of Florida,
P.\ O.\ Box 118440, Gainesville, FL 32611--8440}
\date{18 July 1996}
\maketitle

\begin{abstract}
Perturbative scaling is applied to the Anderson model for a localized
level coupled to a Fermi system in which the density of states varies like
$|\epsilon|^r$ near the Fermi energy ($\epsilon=0$).
This model with $r=1$ or $2$ may describe magnetic impurities in
unconventional superconductors and certain semiconductors.
The pseudo-gap is found to suppress mixed valence in favor of local-moment
behavior.
However, the magnitude of the exchange coupling $J$ in the local-moment regime
is reduced, thereby decreasing the parameter range within which
the impurity becomes Kondo-screened at low temperatures.
\end{abstract}


]
\narrowtext

There exists a class of ``gapless'' Fermi systems which exhibit a pseudo-gap
in the effective density of states $\rho(\epsilon)$ at the Fermi level, taken
to be $\epsilon=0$.
For instance, the valence and conduction bands of certain semiconductors ---
including Pb$_{1-x}$Sn$_x$Te at a critical composition,\cite{Hohler}
and PbTe-SnTe heterojunctions \cite{Volkov} --- touch in such a way that, for
small $|\epsilon|$, $\rho(\epsilon)$ is proportional to $|\epsilon|^{d-1}$
in $d$ spatial dimensions.
The quasiparticle density of states in an unconventional superconductor can
vary like $|\epsilon|$ or $|\epsilon|^2$ near line or point
nodes in the gap.\cite{Sigrist}
Heavy-fermion and cuprate superconductors are candidates for this behavior.
Electrons in a strong magnetic field \cite{Fisher} and exotic phases of the
Hubbard model \cite{Baskaran} are also predicted to exhibit a linear
pseudo-gap in two dimensions.
Finally, the single-particle density of states in the one-dimensional
Luttinger model varies like $|\epsilon|^{2\alpha}$, where $\alpha$ is a
positive number which varies continuously with the strength of the repulsive
Coulomb interactions in the bulk.\cite{luttinger}

Recently there has been considerable interest
\cite{Withoff,Cassanello,Borkowski:92,Chen,KI:NRG,Borkowski:96} in the behavior
of magnetic impurities in gapless systems having a power-law
density of states, $\rho(\epsilon)=\rho_0|\epsilon|^r$.
Poor man's scaling for the spin-$\frac{1}{2}$ (impurity degeneracy $N=2$)
Kondo model \cite{Withoff} and large-$N$ treatments \cite{Withoff,Cassanello}
indicate that a Kondo effect (i.e., complete quenching of a local magnetic
moment in the limit of low temperatures $T$) takes place only if the
antiferromagnetic electron-impurity exchange $\rho_0 J$ exceeds a critical
value, $\rho_0 J_c\approx r$; otherwise, the impurity decouples from the band.
A large-$N$ study of magnetic impurities in gapless superconductors
\cite{Borkowski:92} yields similar results, except that for $r\le 1$ or $N=2$,
any finite impurity concentration drives $J_c$ to zero.
Numerical renormalization-group calculations for the $N=2$ case, both
at\cite{Chen} and near\cite{KI:NRG} particle-hole symmetry, show that
$J_c\rightarrow\infty$ for all $r>\frac{1}{2}$, while for $r<\frac{1}{2}$ the
strong-coupling limit exhibits anomalous properties, including a non-zero
moment.
Away from this symmetry, a finite value of $J_c$, roughly proportional to
$r$, is recovered;
for $J>J_c$ the impurity spin is completely screened at $T=0$, but an
electron phase shift of $\pi$ suggests that the impurity contribution to the
resistivity vanishes, instead of taking its maximal possible value as it does
in the conventional Kondo effect.\cite{KI:NRG}

The Kondo model presupposes the existence of a local moment, i.e., an impurity
level having an average occupancy $\langle n_d\rangle = 1$.
This paper reports the first systematic exploration of local-moment formation
in gapless systems.
Poor man's scaling\cite{Anderson:scaling} is applied to the Anderson
impurity model,\cite{Anderson:model} in which mixed-valence
($0<\langle n_d\rangle<1$) and empty-impurity ($\langle n_d\rangle\approx 0$)
regimes compete with local-moment behavior.
Concentrating on the case of a localized level which lies within a
power-law pseudo-gap, we show that the reduction in the density of states near
the Fermi level has three main effects, each of which grows more pronounced as
$r$ increases:
(1) The mixed-valence region of parameter space shrinks, and for $r\ge 1$
disappears altogether.
(2) The local-moment regime expands.
(3) The value of the Kondo $J$ on entry to the local-moment regime is reduced.
Since the threshold $J$ for a Kondo effect rises with $r$ (see above),
these results imply --- at least in the cases of greatest interest, $r=1$
and $2$ --- that there is a large region of phase space in which the
low-temperature state has an uncompensated local moment.
This should be contrasted with systems having a regular density of states, in
which an Anderson impurity is always quenched at zero temperature.

We start with the Anderson model,\cite{Anderson:model} written in
one-dimensional form:
\begin{eqnarray}
  H & = & \sum_{\sigma=\uparrow,\downarrow}
	  \int_{-D}^D \! d \epsilon \; \epsilon
	  c^\dagger_{\epsilon \sigma} c^{\rule{0ex}{1.35ex}}_{\epsilon \sigma}
          + \epsilon_d n_d + U n_{d \uparrow} n_{d \downarrow} \nonumber \\
    &   & + \sum_{\sigma=\uparrow,\downarrow}
	  \int_{-D}^D \! d \epsilon \; V \sqrt{\rho (\epsilon)} 
          \; ( c^\dagger_{\epsilon \sigma} d^{\rule{0ex}{1.35ex}}_{\sigma} +
          \text{ h.c.} ) .
							\label{H_1d}
\end{eqnarray}
The conduction band is taken to be isotropic in momentum space and to extend
in energy over a range $\pm D$ about the Fermi energy;
the operators $c^{\rule{0ex}{1.35ex}}_{\epsilon\sigma}$ are normalized such
that
$\{ c^\dagger_{\epsilon \sigma} ,
    c^{\rule{0ex}{1.35ex}}_{\epsilon' \sigma'} \} =
       \delta (\epsilon - \epsilon') \delta_{\sigma,\sigma'}$.
The localized level is described by its energy $\epsilon_d$ relative to the
Fermi level, and the cost in Coulomb energy, $U>0$, when it is doubly
occupied.
We have assumed the hybridization between the band and the impurity is purely
local; $V$ can be taken to be a positive real number.
The interesting physics of this model occurs when $V$ and $|\epsilon_d|$
are smaller than $U$ and $D$.

{\em Pure power-law density of states.}
We first consider the density of states introduced in
Ref.~\onlinecite{Withoff}:
\begin{equation}
  \rho (\epsilon) = \left \{ \begin{array}{ll}
                       \rho_0 \left | \epsilon / D \right |^{r} , \quad
			   & |\epsilon| \le D ; \\[0.5ex]
                        0, & \text{otherwise} . \end{array} \right. 
							\label{pure-DOS}
\end{equation}
The exponent $r$ can take any non-negative value, with $r=0$ representing a
constant density of states;
$\rho_0$ is chosen so that $\int_{-D}^D\rho(\epsilon) d\epsilon = 1$.
(Later in the paper we will examine a more realistic case, in which the
power-law variation is restricted to the vicinity of the Fermi level.)

In order to understand the behavior of Eq.~(\ref{H_1d}) at low temperatures
$T$, we apply poor man's scaling.\cite{Anderson:scaling}
In this approach, electronic states with energies $|\epsilon|\gg T$
are progressively integrated out, yielding an effective description of the
problem in terms of fewer degrees of freedom.
Consider first an incremental reduction of the bandwidth from $D$ to
$D'\equiv D(1+\delta \ln D)<D$.
The aim is to represent the same physical system by an effective Hamiltonian
of the form of Eq.~(\ref{H_1d}) and a density of states given by
Eq.~(\ref{pure-DOS}), but with $D$ replaced by $D'$ in both equations.
This requires the couplings entering Eq.~(\ref{H_1d}) to be adjusted to
account for the states which have been eliminated.
Provided that $\delta D$, $\epsilon_d$, and $V$ remain small compared to $D$,
these renormalizations can be computed within perturbation theory.
The band reduction can then be iterated, leading to differential equations
for the couplings as functions of the effective bandwidth.

Poor man's scaling neglects higher-order corrections which conceivably
could accumulate to become important at low energies.
However, the scaling picture presented below is supported, both qualitatively
and quantitatively, by non-perturbative renormalization-group calculations.
\cite{KI:unpub}
Details of the non-perturbative treatment lie beyond the scope of the
present paper.

In previous implementations of poor man's scaling for the Anderson model with
a constant density of states,\cite{Jefferson,Haldane} it was found that
eliminating all states with energies $D' < |\epsilon| \le D$ produces a
leading correction
\begin{equation}
   \delta \epsilon_d = -\rho(D) V^2 \delta \ln D
			 = -(\Gamma/\pi) \, \delta \ln D ,
							\label{delta-Ed}
\end{equation}
where $\Gamma = \pi \rho_0 V^2$.
The corrections to $U$ and $V$ enter at higher order in
small couplings, and can be neglected.

For a power-law density of states, an additional correction is required
because the coupling entering Eq.~(\ref{H_1d}) is not $V$, but
$\sqrt{\rho(\epsilon)}V\equiv\sqrt{\Gamma|\epsilon/D|^r}$.
Replacing $D$ by $D'$ in Eq.~(\ref{pure-DOS}) increases $\rho(\epsilon)$
by a factor of $(D/D')^r$.
$\Gamma$ must be reduced by the same factor so that the physical coupling
remains unaffected by the change of variable, i.e., we require
$\Gamma\,|\,\epsilon/D|^r = (\Gamma+\delta\Gamma)\,|\,\epsilon/D'|^r$,
which gives
\begin{equation}
   \delta \Gamma = r \Gamma \delta \ln D .
							\label{delta-Gamma}
\end{equation}
The novel behaviors of an Anderson impurity in gapless systems all stem from
this correction to $\Gamma$, which has no counterpart for a constant density
of states.

\begin{figure}[t]
\centerline{
\vbox{\epsfxsize=75mm \epsfbox{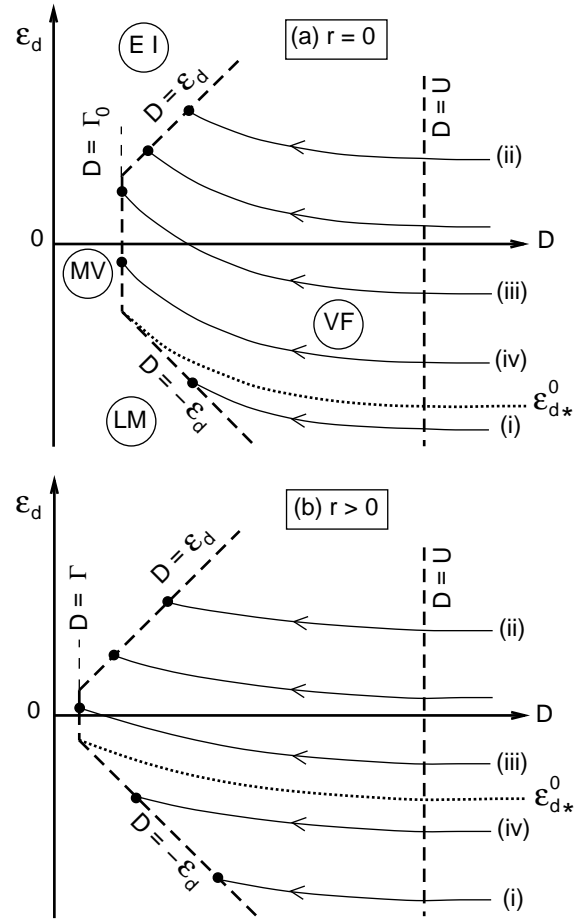}}
}
\vspace{2ex}
\caption{
Scaling of the impurity energy $\epsilon_d$ with the bandwidth $D$,
shown schematically for (a) $r=0$; (b) $0<r<1$.
Renormalization of $\epsilon_d$ begins on entry into the valence-fluctuation
(VF) regime, at $D\approx U$.
Each trajectory ends at a crossover to local-moment
(LM), empty-impurity (EI) or mixed-valence (MV) behavior.
Note that the range of bare impurity energies that scale to the LM regime
($\epsilon_d^0<\epsilon_{d*}^0$) is much greater in (b) than in (a).
}
\label{fig:scaling}
\end{figure}

Equations (\ref{delta-Ed}) and~(\ref{delta-Gamma}) can be integrated to give
the effective couplings $\epsilon_d(D)$ and $\Gamma(D)$ at an arbitrary $D$
in terms of the initial bandwidth $D_0$ and the bare couplings
$\epsilon_d^0\equiv\epsilon_d(D_0)$ and $\Gamma_0\equiv\Gamma(D_0)$:
\begin{eqnarray}
  \Gamma(D) &=& \Gamma_0 \cdot ( D / D_0 )^r ,		\label{Gamma} \\[0.5ex]
  \epsilon_d(D) &=& \epsilon_d^0  + \frac{\Gamma_0}{\pi r}
				    \left[ u^r - ( D / D_0 )^r \right ] .
							\label{Ed}
\end{eqnarray}
[The quantity $u\!\equiv\!\text{min}(1,U/D_0)$ appears because, strictly
speaking, Eq.~(\ref{delta-Ed}) applies only in the range $D\lesssim U$;
the scaling of $\epsilon_d$ is negligible for $D\gtrsim U$.]
For a constant density of states, by contrast, only $\epsilon_d$
renormalizes \cite{Haldane}:
\begin{equation}
   \epsilon_d(D)  =  \epsilon_d^0  + (\Gamma_0 / \pi) \ln ( u D_0 / D ) 
	\qquad (r=0).
							\label{Ed_r=0}
\end{equation}
It should be emphasized that $U$ does not renormalize significantly for
any $r$.

It is instructive to compare the behavior of systems with $r=0$ and $r>0$
as the bandwidth $D$ is progressively reduced.
We assume that the bare parameters are such that $|\epsilon_d^0|,
\Gamma_0 < U < D_0$.
Initially, all four impurity configurations are active, and the impurity
susceptibility $\chi_{\text{\it imp}}$ satisfies \cite{units}
$T\chi_{\text{\it imp}}\approx 1/8$.

Once $D$ is scaled below $U$, the doubly occupied impurity state becomes
frozen out, and the system enters the {\em valence-fluctuation\/} regime.
Real charge fluctuations between the remaining impurity states lead to a
susceptibility $T\chi_{\text{\it imp}}=1/6$.
Here, $\epsilon_d$ scales with $D$ according to Eq.~(\ref{Ed})
or Eq.~(\ref{Ed_r=0}).
This renormalization is represented schematically in Fig.~\ref{fig:scaling},
which shows trajectories $\epsilon_d$ vs.\ $D$, starting from different bare
values.
(The value of $\Gamma_0$ is taken to be the same for all curves.)
Each trajectory in Fig.~\ref{fig:scaling} terminates when either
$\Gamma$ or $|\epsilon_d|$ grows to equal $D$.
At this point perturbation theory breaks down, and there is a crossover to one
of three regimes in which real charge fluctuations are frozen out:

{\em Local moment:}
If $\epsilon_d$ becomes sufficiently large and negative, the impurity acquires
a spin, and the susceptibility rises to $T\chi_{\text{\it imp}}\approx 1/4$.
The scale for this crossover is a solution of the equation
$D_{\text{\it LM}} \equiv -\epsilon_d(D_{\text{\it LM}})$;
see curves (i) in Fig.~\ref{fig:scaling}.

{\em Empty impurity:}
If, instead, $\epsilon_d$ becomes equal to $+D$ [e.g., curves (ii) in
Fig.~\ref{fig:scaling}], the localized level is completely depopulated
and $T\chi_{\text{\it imp}}$ drops rapidly to zero.

{\em Mixed valence:}
Finally, $\Gamma/D$ may become of order unity,
at a bandwidth $D_{\text{\it MV}}\equiv\Gamma(D_{\text{\it MV}})$;
see curves (iii) in Fig.~\ref{fig:scaling}.
In this case, $T\chi_{\text{\it imp}}\rightarrow 0$, but even at $D=0$ the
value of $\langle n_d\rangle$ differs significantly from both 0 and 1.

There are a number of notable differences between the cases $r=0$ and $r>0$
shown in Fig.~\ref{fig:scaling}:

(1) The trajectories in the valence-fluctuation regime are flatter for
the system with a pseudo-gap.
The larger the value of $r$, the more the renormalization of
$\epsilon_d$ is inhibited by the decrease in $\Gamma$.
Consider, for example, the maximum possible shift in the impurity
energy: according to Eq.~(\ref{Ed}),
$\epsilon_d(0)-\epsilon_d^0=(\Gamma_0 u^r)/(\pi r)$,
whereas for $r=0$ the shift in $\epsilon_d$ is unbounded.
For $r\ge\frac{1}{2}$ (say) and $\Gamma_0\ll|\epsilon_d^0|$, it is a
reasonable first approximation to neglect the renormalization
of $\epsilon_d$ altogether.

(2) As $r$ increases from zero, the crossover scale for the mixed-valence
regime is pushed down:
\begin{equation}
  D_{\text{\it MV}} = \left \{ \begin{array}{ll}
		      \Gamma_0 \cdot (\Gamma_0 / D_0)^{r / (1-r)}, \quad
							& 0\le r<1; \\
		      0 ,                               & r \ge 1 .
		      \end{array} \right. 
							\label{D_MV}
\end{equation}
For $r\ge 1$, the ratio $\Gamma/D$ always {\em decreases\/} under scaling,
which completely rules out mixed-valence behavior;
instead, the system must eventually enter either the local-moment
regime or the empty-impurity regime.

\begin{figure}
\centerline{
\vbox{\epsfxsize=80mm \epsfbox{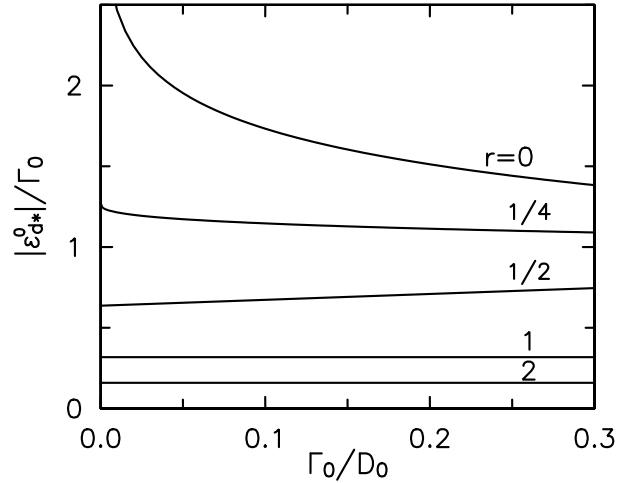}}
}
\vspace{2ex}
\caption{
Boundary of the local-moment regime, $|\epsilon_{d*}^0|/\Gamma_0$, plotted
vs.\ $\Gamma_0$ for different values of $r$.
Any bare impurity energy $\epsilon_d^0 < \epsilon_{d*}^0$
eventually leads to local-moment behavior.
}
\label{fig:Ed_max}
\end{figure}

(3) The depression of the mixed-valence scale $D_{\text{\it MV}}$ and the
flattening of the scaling trajectories both tend to widen the range of
bare impurity energies which eventually result in local-moment behavior.
Curves (iv) in Fig.~\ref{fig:scaling} show a case in which the pseudo-gap
diverts a trajectory away from the mixed-valence regime to intersect the
local-moment line.
To quantify this trend, let $\epsilon_{d*}^0$ be the largest
(least negative) bare impurity energy which flows to the local-moment regime
(see Fig.~\ref{fig:scaling}).
This energy is given implicitly by the equation $D_{\text{\it MV}} =
D_{\text{\it LM}}$.
Using Eqs.~(\ref{Ed}) and~(\ref{D_MV}), one obtains
\begin{equation}
   \epsilon_{d*}^0
	= \left( \frac{1}{\pi r} - 1 \right) D_{\text{\it MV}}
          - \frac{\Gamma_0 u^r}{\pi r} ,
							\label{Ed*}
\end{equation}
The plot of $\epsilon_{d*}^0/\Gamma_0$ in Fig.~\ref{fig:Ed_max} clearly shows
the expansion of the local-moment region with increasing $r$.

\begin{figure}
\centerline{
\vbox{\epsfxsize=80mm \epsfbox{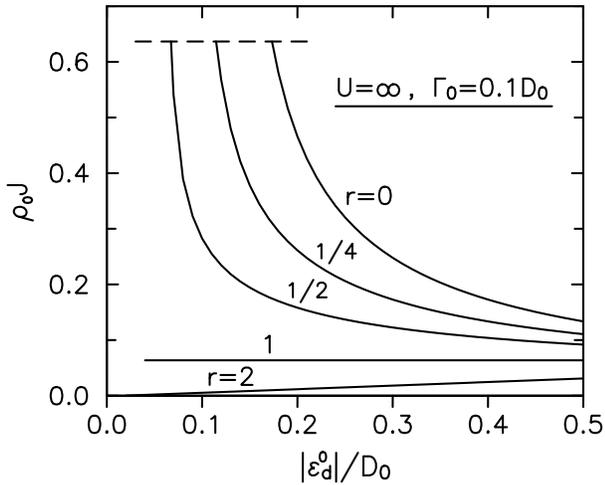}}
}
\vspace{2ex}
\caption{
Exchange coupling $\rho_0 J$ on entry to the local-moment regime, plotted
vs.\ $|\epsilon_d^0|$ for different values of $r$.
Notice that the exchange coupling decreases as $r$ increases.
}
\label{fig:J}
\end{figure}

(4) At the point where a scaling trajectory enters the local-moment regime,
the Anderson model can be mapped via a Schrieffer-Wolff transformation
\cite{Schrieffer} onto the Kondo model with an effective exchange coupling
\begin{eqnarray}
\rho_0 J & = & \frac{2 \Gamma}{\pi |\epsilon_d|}
                  +\frac{2 \Gamma}{\pi (U + \epsilon_d)}
							\nonumber \\
         & = & \left[ \frac{2 \Gamma_0}{\pi D_{\text{\it LM}}}
                     +\frac{2 \Gamma_0}{\pi (U-D_{\text{\it LM}})} \right]
	       \left( \frac{D_{\text{\it LM}}}{D_{0}} \right)^r.
							\label{SW_transf}
\end{eqnarray}
For given impurity parameters ($\epsilon_d^0$, $U$, and $\Gamma_0$),
the Kondo $J$ for $r>0$ is reduced compared to that for a regular density
of states, due both to the depression of $\Gamma$ and to the weaker
renormalization of $\epsilon_d$.
A lower bound on the reduction factor, obtained by neglecting
the renormalization of $\epsilon_d$, is $|\epsilon_d^0/D_0|^r$.
This effect is illustrated in Fig.~\ref{fig:J}, which plots $\rho_0 J$ as a
function of $\epsilon_d^0$, for $U=\infty$, $\Gamma_0=D_0/10$, and several
values of $r$.
For $r<1$, $\rho_0 J$ rises to reach $2/\pi$ (the dashed line in
Fig.~\ref{fig:J}) at $\epsilon_d^0=\epsilon_{d*}^0$, the boundary of
the local-moment region, whereas for $r>1$, $\rho_0 J$ decreases instead.
Note that only for $r=0$ and $r=1/4$ is the condition $\rho_0 J\gtrsim r$
for the existence of a Kondo effect satisfied over any extended range of
$\epsilon_d^0$.
This observation extends to other values of $\Gamma_0$ and $U$.

{\em Restricted power-law density of states:}
In most gapless systems, the power-law variation of the density of states
does not extend over the entire band in the manner assumed in
Eq.~(\ref{pure-DOS}).
We therefore repeat the preceding analysis for a more realistic density of
states which rolls over to a constant beyond a region of width $\pm\Delta$
about the Fermi level, i.e., $\rho(\epsilon) = \rho_0 |\epsilon/\Delta|^r$
for $|\epsilon|<\Delta$, but $\rho(\epsilon)= \rho_0$
for $\Delta < | \epsilon | \le D$.

At energies much greater than $\Delta$, scaling should proceed very much
as for a constant density of states.
It is quite possible for the system to pass out of the valence-fluctuation
regime before the pseudo-gap can have any real effect.
However, we are more interested in values of $\epsilon_d^0$ and $\Gamma_0$
which are sufficiently small that the bandwidth can be scaled into the range
$D<\Delta$, where Eqs.~(\ref{delta-Ed}) and~(\ref{delta-Gamma}) must apply.
The subsequent renormalization of $\Gamma$ and $\epsilon_d$ is identical
to that for a system having a pure power-law density of states, but with 
a bare bandwidth $\Delta$, and a bare impurity energy $\epsilon_d(\Delta)$
calculated from Eq.~(\ref{Ed_r=0}).
The qualitative effects of the pseudo-gap should therefore be the same as
those found above, although the magnitude of these effects will
certainly decrease as the width of the pseudo-gap becomes smaller.
In particular, there will remain an expansion of the local moment regime,
in which the Kondo $J$ will still be reduced relative to the case $r=0$
by a factor of at least $|\epsilon_d(\Delta)/\Delta|^r$.

Finally, we note a parallel between our findings for an impurity in a
non-interacting background and results showing that an Anderson impurity
in a Luttinger liquid has an expanded local-moment regime.\cite{Phillips}
This similarity appears surprising, since in an interacting
one-dimensional electron gas, independent fermionic excitations are
replaced by collective bosonic modes.\cite{luttinger}
It seems, though, that bulk interactions affect the valence-fluctuation
regime of the impurity only by generating a power-law spectrum of
one-electron states which hybridize with the localized level.
The similarity with the non-interacting case does not extend
into the local-moment regime: there is no threshold value of $J$
required to achieve a Kondo effect in a Luttinger liquid.\cite{Lee}

In summary, we have investigated local-moment formation in Fermi systems
having a density of states which vanishes as $|\epsilon|^r$ near the
Fermi energy.
The pseudo-gap strongly suppresses hybridization between conduction electrons
and an impurity level, and weakens the renormalization of the impurity energy.
These effects in turn expand the range of bare impurity parameters that lead
to localization of a spin at the impurity site, but reduce the value of the
Kondo exchange coupling of this spin to the band, and thus decrease
the likelihood that the moment will be Kondo-screened at low temperatures.

We thank P.\ Hirschfeld for useful comments on the manuscript.
Portions of this work were carried out while one of us (K.I.) was visiting the
Institute for Theoretical Physics, Santa Barbara.
This research was supported in part by NSF Grants No.\ DMR93--16587
and PHY94--07194.



\begin{thebibliography}{9}
\bibitem[*]{byline}
    Affiliated with the National High Magnetic Field Laboratory,
    Tallahassee, Florida 32306.
\bibitem{Hohler}
    {\it Narrow-Gap Semiconductors,} Springer Tracts in Modern Physics,
    Vol.\ 98, ed.\ G.\ H\"{o}hler (Springer-Verlag, Berlin, 1983).
\bibitem{Volkov}
    B.\ A.\ Volkov and O.\ A.\ Pankratov,
    Pis'ma Zh.\ Eksp.\ Teor.\ Fiz.\ {\bf 42}, 145 (1985)
    [JETP Lett.\ {\bf 42}, 178 (1985)].
\bibitem{Sigrist}
    M.\ Sigrist and K.\ Ueda,
    Rev.\ Mod.\ Phys.\ {\bf 63}, 239 (1991).
\bibitem{Fisher}
    M.\ P.\ A Fisher and E.\ Fradkin,
    Nucl.\ Phys.\ {\bf B251} [FS13], 457 (1985).
\bibitem{Baskaran}
    G.\ Baskaran, Z.\ Zou, and P.\ W.\ Anderson,
    Solid State Commun.\ {\bf 63}, 973 (1987);
    A.\ Ruckenstein, P.\ J.\ Hirschfeld, and J.\ Appel,
    Phys.\ Rev.\ {\bf B36}, 857 (1987);
    I.\ Affleck and J.\ B.\ Marston, {\it ibid.} {\bf 37}, 3774 (1988);
    G.\ Kotliar, {\it ibid.} {\bf 37}, 3664 (1988).
\bibitem{luttinger}
    For a recent review, see: J.\ Voit, to appear in Rep.\ Prog.\ Phys.\
    (cond-mat/9510014).
\bibitem{Withoff}
    D.\ Withoff and E.\ Fradkin, Phys.\ Rev.\ Lett.\ {\bf 64}, 1835 (1990).
\bibitem{Cassanello}
    C.\ R.\ Cassanello and E.\ Fradkin, 1996 preprint (cond-mat/9512064).
\bibitem{Borkowski:92}
    L.\ S.\ Borkowski and P.\ J.\ Hirschfeld,
    Phys.\ Rev.\ {\bf B46}, 9274 (1992);
    J.\ Low.\ Temp.\ Phys.\ {\bf 96}, 185 (1994).
\bibitem{Chen}
    K.\ Chen and C.\ Jayaprakash,
    J.\ Phys.: Condens.\ Matter {\bf 7}, L491 (1995).
\bibitem{KI:NRG}
    K.\ Ingersent, to appear in Phys.\ Rev.\ B (cond-mat/9605025).
\bibitem{Borkowski:96}
    L.\ S.\ Borkowski, preprint (cond-mat/9606075).
\bibitem{Anderson:scaling}
    P.\ W.\ Anderson, J.\ Phys.\ C {\bf 3}, 2436 (1970).
\bibitem{Anderson:model}
    P.\ W.\ Anderson, Phys.\ Rev.\ {\bf 124}, 41 (1961).
\bibitem{KI:unpub}
    K.\ Ingersent, unpublished.
\bibitem{Jefferson}
    J.\ H.\ Jefferson, J.\ Phys.\ C {\bf 10}, 3589 (1977).
\bibitem{Haldane}
    F.\ D.\ M.\ Haldane, Phys.\ Rev.\ Lett.\ {\bf 40}, 416, 911(E) (1978).
\bibitem{units}
    The units are chosen such that $k_B=g\mu_B=1$, where $g$ is the
	electronic $g$-factor and $\mu_B$ is the Bohr magneton.
\bibitem{Schrieffer}
    J.\ R.\ Schrieffer and P.\ A.\ Wolff, Phys.\ Rev.\ {\bf 149}, 491 (1966).
\bibitem{Phillips}
    P.\ Phillips and N.\ Sandler, Phys.\ Rev.\ B {\bf 53}, 468 (1996);
    A.\ Schiller and K.\ Ingersent, in preparation.
\bibitem{Lee}
    D.-H.\ Lee and J.\ Toner, Phys.\ Rev.\ Lett.\ {\bf 69}, 3378 (1992);
    A. Furusaki and N. Nagaosa, {\it ibid.} {\bf 72}, 892 (1994).
\end{thebibliography}
\end{document}